# Neural dynamics of emotion and cognition:
# from trajectories to underlying neural geometry

Luiz Pessoa
pessoa@umd.edu

Department of Psychology

Department of Electrical and Computer Engineering

Maryland Neuroimaging Center

University of Maryland, College Park, USA

**Abstract**

This paper describes the outlines of a research program for understanding the cognitive-emotional brain, with an emphasis on the issue of *dynamics*: How can we study, characterize, and understand the neural underpinnings of cognitive-emotional behaviors as inherently dynamic processes? The framework embraces many of the central themes developed by Steve Grossberg in his extensive body of work in the past 50 years. By embracing head on the leitmotifs of dynamics, decentralized computation, emergence, selection and competition, and autonomy, it is proposed that a science of the mind-brain can be developed that is built upon a solid foundation of understanding behavior while employing computational and mathematical tools in an integral manner. A key implication of the framework is that standard ways of thinking about causation are inadequate when unravelling the workings of a complex system such as the brain. Instead, it is proposed that researchers should focus on determining the dynamic multivariate structure of brain data. Accordingly, central problems become to characterize the dimensionality of neural trajectories, and the geometry of the underlying neural space. At a time when the development of neurotechniques has reached a fever pitch, neuroscience needs to redirect its focus and invest comparable energy in the conceptual and theoretical dimensions of its research endeavor. Otherwise we run the risk of being able to measure "every atom" in the brain in a theoretical vacuum.

## 1. The problem of emotion, cognition, and behavior

This paper describes the outlines of a research program for understanding the cognitive-emotional brain, with an emphasis on *dynamics*: How can we study, characterize, and understand the neural underpinnings of cognitive-emotional behaviors as inherently dynamic processes?

At the outset, I propose eliminating the distinction between emotion and cognition (Pessoa, 2018a). What is emotion? What are its defining characteristics? Are emotions distinct from feelings? Researchers have debated, and in fact agonized over, such questions for a very long time. And the debate continues. For example, nine essays are dedicated to the topic in the latest edition of *The Nature of Emotion* (Fox, Lapate, Davidson, & Shackman, 2018); and additional suggestions continue appearing (see, Fox, 2018). Such pursuit of the "essence of emotion" appears misguided. What researchers of the mind and brain are interested in, it could be argued, is understanding behaviors. Mind scientists seek to understand the structure of behaviors, their inherent logic. Brain scientists strive to unravel how the two domains, mental and neural, map to one another during behaviors.

The framework described here is strongly influenced by many lines of research and thinking (as any intellectual endeavor, of course), and most of all by the research by Steve Grossberg[1].

## 2. Grossbergian themes

Grossberg has developed his theoretical framework for over 50 years. The breadth of his thinking is so enormous as to defy understanding. In this section, I will describe a series of themes that permeate his work, sometimes very explicitly, at times less so. Although the remainder of the paper will build upon more directly on only a few of the themes – and centrally on dynamics – all of them are viewed as *essential* to building an understanding of the cognitive-emotional brain.

---

[1] I was a graduate student at the Department of Cognitive and Neural Systems from 1990 to 1995, and worked closely with Steve during the last year of my PhD. This work was continued after I returned to Brazil until 1998. Steve also taught an enormously inspiring informal seminar during my second or third year in which he outlined his research program. His infinite energy and untiring guidance have been constant sources of inspiration in my career.

*2.1 Dynamics*

This theme is so central to Grossberg's work that it is fair to say that without it the work would not exist. In Grossberg's very first publication[2], he states:

> Fundamental to the motivation of the new theory is the realization that the *dynamics* of many psychological problems may be viewed from a unified point of view once the geometrical substrates that characterize each separate problem are elaborated and distinguished (Grossberg, 1964; italics added).

The very first equation of his opus (Grossberg, 1964) reads as follows:

$$\frac{ds_k}{dt} = \alpha(M - s_k)T_k - D_k,$$

where that the "activation" $s_k$ was defined via a grow process whereby $s_k$ increased toward $M$ at rate $\alpha$ and its total input $T_k$ (itself dependent on other activations), while also subject to a simple exponential decay, $D_k$.

At first, it would appear that one would hardly have to emphasize dynamics as an important principle. Yet, experimental brain research is frequently, and even preponderantly, quasi-static. Data from almost any measurement modality (physiology, functional MRI, etc.) are epoched in terms of trials or segments that largely discard most temporal information.

*2.2 Behavior*

Consideration of a very extensive body of behavioral data is essential. Contrast this to a sort of "tunnel vision" that is unfortunately widespread, as all too often researchers break into cliques that focus on apparently distinct sets of phenomena. For example, research addressing "appetitive" and "aversive" processing has been carried out by largely separate communities. More generally, researchers focus on "motivation" *or* "emotion," on "cognition" *or* "emotion," and so on. But behaviors do not obey boundaries, and thinking about diverse sources of data is necessary for deeper understanding.

---

[2] A monograph published while in graduate school with over 400 pages.

Behavior is the founding pillar for explaining the brain. This reads like a truism. Unfortunately, it is not, as suggested for example by the popularity of the recent paper by Krakauer and colleagues (2017) in the powerful journal *Neuron* (see also Gomez-Marin, 2016; Gomez-Marin et al., 2014). Their general call to arms to embrace behavior and eschew a neuronal reductionistic bias has resonated with those who believe that ever more sophisticated measurement techniques are not enough to dissect the brain.

What if the activity of every neuron could recorded in the brain of an animal during a certain behavior (see Ahrens et al., 2012; Lovett-Barron et al., 2017). What would be gained by doing so? Consider a device that can measure the exact state of a modern Airbus 380 aircraft (which weighs more than a million pounds), say an image of every atom (aircraft are mostly made of aluminum) at millisecond resolution. An adequate level of description of the aircraft and its parts is in terms of fluid dynamics and related aerodynamics, where issues related to compressible flow, turbulence, and boundary layers are important. Therefore, although it is conceivable that this future device could provide some useful information, the point made here is that additional data are only minimally useful without more advanced theoretical understanding – of both mind (that is behavior) and brain.

Returning to the theme of dynamics, in parallel with the way data are analyzed, behavior is frequently conceptualized in terms of discrete trials of relatively short duration. This approach is understandable from the perspective of experimental scientists who need trial averaging to handle noise. But behavior itself is inherently temporal, and neglecting that aspect seriously limits research progress.

*2.3 Decentralization, heterarchy*

Understanding systems in terms of the *interactions* between their parts fosters a way of thinking that favors decentralized organization. It is the coordination between the multiple parts that leads to the behaviors of interest, not a "controller" that dictates the function of the system. In many sophisticated systems, and the brain is no exception, it is natural to think that many of its chief functions depend on centralized processes. For example, the prefrontal cortex may be viewed as a uniquely positioned brain sector where multiple types of information converge, allowing it to then guide behavior (Fuster, 2000; Miller and Cohen, 2001). A contrasting view is one in which processing takes place in a *distributed* fashion via the interactions of constituent parts. Accordingly, instead of information flowing hierarchically to an "apex region" where all the pieces are combined, information flows in multiple directions without a strict hierarchy. An organization of this sort is termed a *heterarchy* to emphasize the notion that the flow of information is multidirectional (McCulloch, 1945).

A vivid illustration of the problem of centralization involves "executive control" processes. Early models of executive function were built around the notion of a "controller" – essentially a homunculus – that regulates lower-level systems when needed (e.g., Baddeley, 1996; see Shallice, 1988). The inherent problems with such an approach were eventually recognized by several investigators, who called for a "fractionation" of the executive in more manageable (that is, less intelligent) units (Monsell and Driver, 2000). Functions such as "shifting," "updating," and "inhibition" (Miyake et al., 2000) became more prevalent when describing the executive. However, time and again the use of such constructs has amounted to a way of redescribing the object of study rather than actual explanation. To this day, the goal of "banishing the homunculus" remains a formidable challenge (Verbruggen et al., 2014).

The historical conceptualization of the hypothalamus provides another useful example (for an excellent discussion, see Morgane, 1979). This structure is generally referred to as the "head ganglion of the autonomic nervous system." This rubric encapsulates a hierarchical theoretical view based on the idea of "descending" control: the area functions as a central controller of structures along the extent of the brainstem. Indeed, the hypothalamus has robust projections to multiple brainstem sites. However, no area is simply an outflow region (and thus a "head"); all areas receive multiple inputs. In the case of the hypothalamus, multiple brainstem sites that receive projections from the hypothalamus project back to it, following the general tendency of connections to be bidirectional. More critically, the hypothalamus is extensively and bidirectionally connected with most sectors of the cortex (Nieuwenhuys, Voogd, and Huijzen, 2008; Pessoa, 2017a). Far from a master controller, the hypothalamus is an integral node of cortical-subcortical communication.

*2.4 Emergence*

This is a central concept in all of Grossberg's work, as captured by this recent statement:

> Brain circuits give rise to these distinct psychological functions as *emergent properties* that arise from interactions among brain regions that work together as *functional systems*. (Grossberg, 2018, p. 2; italics in the original).

Von Bertalanffy (1950, p. 135), one of the chief early proponents of complex systems theory, famously asserted "the necessity of investigating not only parts but also relations of organization resulting from a dynamic interaction and manifesting themselves by the difference in behavior of parts in isolation and in the whole organism". But what does it mean to say "difference in behavior of parts in isolation and in the whole organism"? Hence *emergence*[3], a term originally coined in the 1870s to describe instances in chemistry and physiology where new and unpredictable properties appear that are not clearly ascribable to the elements from which they arise.

But what does emergence mean? At the most basic level it reflects the notion that "something new appears." While fascinating, this proposition sits uncomfortably with experimental scientists. As presciently stated by Von Bertalanffy (1950, p. 142) himself, the "exact scientist therefore is inclined to look at these conceptions with justified mistrust." Unfortunately, the picture has not appreciably changed, despite stunning developments in mathematics and physics in understanding nonlinear dynamical systems in the last 50 years. Today, emergence can be defined precisely, and in ways that leave no room for vague allusions to "wholeness" or "system properties." In the present context, the body of work by Grossberg provides a clear demonstration of how "emergent properties" can be precisely defined.

*2.5 Selection and competition*

Selection of information for further analysis is a key problem that needs to be solved for effective behavior. Indeed:

---

[3] The term emergence appears to have been first proposed in the 1870s when used by George Henry Lewes in his book *Problems of Life and Mind* and taken up by Wilhelm Wundt in his *Introduction to Psychology*.

> How can a limited-capacity information processing system that receives a constant stream of diverse inputs be designed to selectively process those inputs that are most significant to the objectives of the system? (Grossberg and Levine, 1987, p. 5015)

If selection is the ubiquitous problem that must be effectively solved, *competition* is the mechanism by which stimuli, objects, actions, and so forth are selected.

*2.6 Autonomy*

Central to Grossberg's theoretical framework is the notion of autonomy:

> *Brains look the way that they do because they embody computational designs whereby individuals autonomously adapt to changing environments in real time.* Grossberg (2018, p. 4; italics in the original).

To understand the cognitive-emotional brain, it is necessary to consider that all animals need to function independently in diverse and challenging conditions and environments – they need to be autonomous.

All vertebrates have a brain architecture that allows a considerable amount of communication and integration of signals (Pessoa, 2018b; Pessoa et al., in preparation). Why this kind of architecture? One possibility is that it confers a high degree of flexibility that allows animals to cope with the complex interactions in their changing habitats, involving predators, prey, potential mates, and so on. Survival may benefit from circuits that can form in a *combinatorial* fashion, as the number of conditions related to the internal and externals worlds of the animal are exceedingly high.

Consider a key system for both appetitive and defensive behaviors, the superior colliculus in the midbrain (Dean et al., 1989; Peek & Card, 2016; Pereira & Moita, 2016). It receives retinal inputs and has outputs that give it access to movements of head and neck, for example. In rodents, the superior colliculus could be involved in implementing the following rule: **If unexpected movement is overhead, flee; otherwise, if movement is in the lower field, consider further exploration.** However, simple rules based on stimulus features do not capture the flexibility of rodent behavior (think how hard it is to catch a rat! see Dean et al., 1989). In particular, rats freeze more frequently to novel stimuli in unfamiliar environments, such as an open field. Clearly, the context in which a stimulus occurs is essential (Peek & Card, 2016; Pereira & Moita, 2016).

More generally, one way to view the more elaborate architecture of birds and mammals (Striedter, 2005) is in terms of the enhanced potential for combinatorial interactions that they afford, such that the manner different signals can influence each other is considerably expanded – and accordingly expand the range of behaviors. This overall type of architecture may produce circuits with local specificity but relatively large-scale sensitivity, a type of *global-within-local* design, which likely contributes to more plastic and sophisticated behaviors. Yet, integration is evolutionarily ancient – it is a hallmark of the vertebrate brain – and could explain the existence of complex behaviors now recognized in all vertebrate taxa.

*2.7 Computational theory*

Brain research is a strongly empirical scientific enterprise. To be sure, research is inspired and guided by conceptual/theoretical thinking, although mostly in a qualitative fashion. But as Rabinovich and colleagues (2006, p. 48) state: "Neural networks [both natural and artificial] are complicated dynamical entities, whose properties are understood only in the simplest cases." Can the complex architecture that supports the cognitive-emotional brain be investigated without formal/mathematical tools? Given the richness of the multi-level interactions, does neuroscience need to migrate to a model that is closer to that of physics? Experimental physicists are not lacking in mathematical sophistication. Neuroscience, in contrast, has evolved into extremely sophisticated "laboratory techniques" that are often divorced from formal approaches. How should we train future generations of brain scientists? Grossberg's position on these questions is easy to predict, as he and colleagues created the Department of Cognitive and Neural Systems at Boston University in 1989 exactly to address this issue.

**3. Decentralized computing: Top-down control versus circuit interactions**

The central claim is that, for interesting behaviors, most of the required explaining is not present at the level of isolated systems (perception, action, etc.) but at the level of the *interactions* between them (Pessoa, 2018a,b). Strictly speaking, however, the concept of an interaction is not opposed to the notion of separable entities, as interactions can also refer to distinct variables, processes, or systems that, themselves, produce effects in a non-additive manner. Accordingly, a better term is *integration*, which implies sufficient intertwining between the putatively separate systems that their individuation becomes a linguistic short-cut.

Let's consider the mechanisms of *fear extinction* (Figure 1A). When a conditioned stimulus (CS) no longer predicts the unconditioned stimulus (UCS) to which it was paired at some point in the past, a new relationship needs to be learned, namely the CS is no longer associated with the UCS – this type of learning is called "extinction." The medial prefrontal cortex (PFC) plays an important role during extinction, as initially revealed via lesioning (Morgan et al., 1993) and subsequently by chemical manipulation of this area (for review, see Dunsmoor et al., 2015). As the medial PFC is extensively interconnected with the amygdala, an early idea was that the former would exert an inhibitory influence on the latter, thereby enabling the extinction of the conditioned response. At this level of description, fear extinction fits the scheme of separate entities interacting to generate a new behavior: cognition (tied to the medial PFC) controlling emotion (tied to the amygdala) in a top-down fashion.

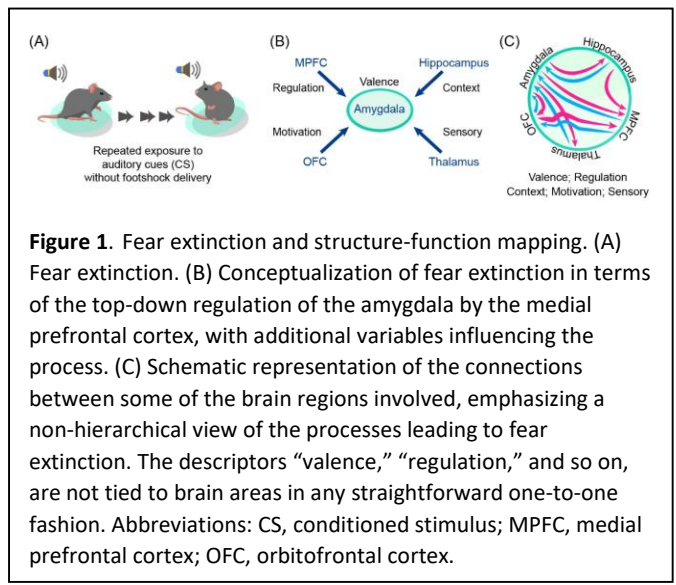

**Figure 1**. Fear extinction and structure-function mapping. (A) Fear extinction. (B) Conceptualization of fear extinction in terms of the top-down regulation of the amygdala by the medial prefrontal cortex, with additional variables influencing the process. (C) Schematic representation of the connections between some of the brain regions involved, emphasizing a non-hierarchical view of the processes leading to fear extinction. The descriptors "valence," "regulation," and so on, are not tied to brain areas in any straightforward one-to-one fashion. Abbreviations: CS, conditioned stimulus; MPFC, medial prefrontal cortex; OFC, orbitofrontal cortex.

Yet, considering the PFC as "top" and the amygdala as "down" does not take into account the richness of the existing neuronal interactions. It is well known that the amygdala plays a critical role in aversive learning, that is, the initial CS-UCS learning. The amygdala plays a critical role in the acquisition and consolidation of fear extinction, too. Chemical blockage of amygdala mechanisms (in the basolateral amygdala) either impair or entirely prevent the acquisition of extinction (Herry et al., 2006). In addition, consolidation of extinction is supported by morphological changes in amygdala synapses (in the basolateral amygdala; see Tovote et al., 2015). These findings, together with the existence of amygdala pathways to the medial PFC, has led some investigators to suggest that the amygdala actually should be viewed as the "top" region in the relationship with the medial PFC (Herry et al., 2008; see also Do-Monte et al., 2015). In fact, multiple cell groups in the amygdala project to the medial PFC, whose outputs in turn influence amygdala signals.

The extinction of conditioned responses is one of the oldest and most widely known findings from psychological science (Dunsmoor et al., 2015). Despite this long history, recent research has greatly expanded our knowledge about this phenomenon, revealing that extinction, far from a simple inhibitory process, is an extremely nuanced learning process. Extinction is now understood to be a form of learning (of the new relationship between the CS and UCS) that itself involves acquisition, retrieval, and consolidation. In other words, it is not a simple inhibitory mechanism of the "fear response" but a sophisticated form of learning. In particular, following extinction, contextual information plays a critical role in determining whether the original fear memory or the new "extinction memory" controls behavior – should the animal fear or not the CS? Accordingly, an elaborate set of neural interactions is needed to support such context sensitivity.

When the CS no longer predicts an aversive outcome, it behooves the animal to take into account that information, such that features of the new environment are learned so as to predict safety. The hippocampus plays a key role in establishing context dependence during extinction learning. There are at least two anatomical routes by which the hippocampus contributes to these processes (Herry et al., 2008; Maren et al., 2013). The first involves direct projections from the hippocampus to the amygdala; the hippocampus is part of a circuit that involves amygdala neurons that are engaged when the behavioral context is *different* from the extinction context (this pathway thus promotes fear). The second, indirect contribution involves dense projections to the medial prefrontal cortex, which appearss to participate in a circuit with the amygdala that indicates safety (this pathway is linked to extinction behaviors).

Another region in the circuit determining if fear should be switched on or off is the thalamus, which is a major player in the processing of biologically significant stimuli (Heimer et al., 2017), as well as a key subcortical–cortical connectivity hub (Pessoa, 2017b). In the past few years, the paraventricular nucleus of the thalamus (PVT) has been established as a thalamic node that interacts with cortico-amygdala circuits for the establishment, retrieval, and maintenance of long-term fear memories (Do-Monte et al., 2015; Penzo et al., 2015). Neurons in the PVT are robustly activated by behaviorally relevant events, including novel ("unfamiliar") stimuli, as well as reinforcing stimuli and their predicting cues (Ren et al., 2018). Notably, PVT responses are influenced by changes in homeostatic state and behavioral context, and inhibition of the PVT suppresses appetitive and aversive learning (Ren et al., 2018). Given that the PVT is bidirectionally connected with the medial PFC, and projects throughout the extended amygdala (central amygdala plus bed nucleus of the stria terminalis), this region is well placed to further refine processing during behavioral conditions eliciting fear extinction.

More generally, during fear extinction – and, in fact, fear acquisition and expression – signals from the amygdala, medial PFC, hippocampus, thalamus, among others, *collectively* determine behavioral responses. These multi-region interactions afford greater behavioral malleability when responding to threat. A more standard approach to attempting to explain fear extinction would be to label each brain region in the following manner, for example: amygdala-valence, medial PFC-regulation, hippocampus-context, thalamus-biological significance, and so on. One could then describe observed behaviors in terms of "standard interactions" (that is, those involving separate entities) between the putative processes (valence, regulation, etc.) (Figure 1B). But if these processes are not separable, they do not encode stable variables that are simply modulated by other variables. In the end, explanations in terms of standard interactions will be found wanting – *integration* is needed (Figure 1C).

## 4. Causation in complex systems

A potentially unappealing aspect of the discussion and conclusion above is that causation is muddied – what causes what during extinction? Dissecting phenomena in terms of their component parts seems like an unimpeachable methodology, to the extent that it can be viewed as almost an axiom of modern science (Deacon, 2011). At the broadest level, the issues at hand speak to how we should study systems as complex as minds and brains.

Understanding causation has always been at the core of the scientific enterprise. To a great extent, the mission of neuroscience is to uncover the nature of the signals observed in different parts of the brain, and to attempt to *disentangle* the potential contributions to those signals. Consider a type of reasoning prevalent in neuroscience, what can be called the *billiard ball* model of causation. In this Newtonian model, force applied to a ball leads to its movement on the table until it hits the target ball (Pessoa, 2017c, 2018c). The reason the target ball moves is obvious; the first ball hits it, and via the force applied to it, it moves. Translated into neural jargon, we can rephrase as follows: a signal external to a brain region excites neurons in that region, which excite or inhibit neurons in another brain region given anatomical pathways connecting them. But this general mode of thinking, which has been very productive in the history of science, is too impoverished when complex systems – the brain for one – are considered.

Mannino and Bressler (2015) highlight two features of the brain that are problematic for standard (Newtonian) causation. First, anatomical connections are frequently bidirectional, leading to bidirectional physiological influences. But if one element causally influences another while the second simultaneously causally influences the first, sometimes called mutual causality (Frankel, 1986), the concept breaks down. Second, *convergence* of anatomical projections implies that multiple regions concurrently influence a single receiving node, making the attribution of unitary causal influences problematic.

Along these lines, in a recent paper I outlined several principles of brain organization that impact the understanding of causality (Pessoa 2017b; see also Pessoa, 2014): 1) massive combinatorial anatomical connectivity; 2) extensive cortical-subcortical anatomical connectional systems; 3) high distributed functional connectivity[4]; 4) overlapping large-scale functional brain networks; and 5) dynamic large-scale functional brain networks. Taken together, the brain basis of emotion-cognition involves distributed, large-scale cortical-subcortical networks. The high degree of signal distribution and integration provides a nexus for the intermixing of information related to perception, cognition, emotion, motivation, and action. Importantly, the functional architecture consists of multiple overlapping networks that are highly dynamic and context-sensitive, such that how a given brain region affiliates with a specific network shifts as a function of task demands and brain state (Najafi et al., 2017. In all, particular cognitive-emotional behaviors can be understood in terms of dynamic *functionally integrated systems,* such as the one involving the amygdala and its cortical-subcortical circuits (Pessoa, 2017b).

---

[4] Functional connectivity refers to the degree of association between time series data of two brain regions, irrespective of their anatomical connectivity status (whether directly connected or not). It is typically estimated based on Pearson's correlation coefficient, although multiple measures of association have been proposed (for example, mutual information).

The upshot is that Newtonian causality provides an extremely poor candidate for explanation in *non-isolable* systems like the brain. The shift advocated here is to move away from individual entities (like billiard balls) and consider the temporal evolution of "multi-particle systems." An example is provided by considering another analogy from physics: the motion of celestial bodies in a gravitational field.

Physicists and mathematicians have been interested in the problem of stability for centuries, and this was a central problem in Newtonian (that is, non-relativistic) celestial mechanics. For example, what types of trajectories do two bodies, such as the earth and the sun, exhibit? The so-called two-body problem was completely solved by Johann Bernoulli in 1734. For more than two bodies (for example, the moon, the earth, and the sun), the problem has vexed mathematicians for centuries. Although the three-body problem cannot be solved in the same sense as the two-body problem, topological properties can be used to classify families of three-body periodic orbits. Figure 2 displays the "yin-yan II" class first reported by

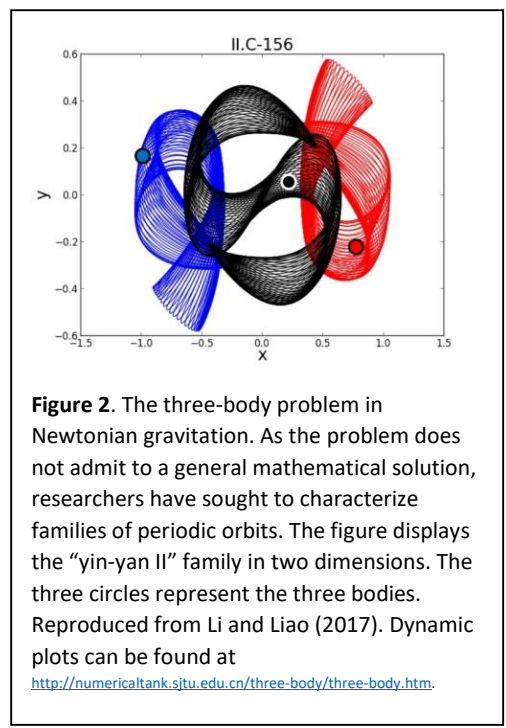

**Figure 2**. The three-body problem in Newtonian gravitation. As the problem does not admit to a general mathematical solution, researchers have sought to characterize families of periodic orbits. The figure displays the "yin-yan II" family in two dimensions. The three circles represent the three bodies. Reproduced from Li and Liao (2017). Dynamic plots can be found at
http://numericaltank.sjtu.edu.cn/three-body/three-body.htm.

Suvakov and Dmitrasinovic (2013) in a breakthrough study describing a large number of new orbit families.

This gravitational three-body problem illustrates the approach to studying complex systems described here: determining and characterizing the temporal evolution of multi-particle systems. The idea will be further developed below after a brief comment on the use of advanced neurotechniques to study causation in neural systems.

4.1 *Optogenetic causation*

In neuroscience, causal efficacy – that is, causal intervention – has become the field's gold standard, and is implicitly or explicitly equated with understanding. Technologies such as optogenetics exist that make it possible to manipulate neural circuits directly and more precisely. However, causal-mechanistic explanations, while valuable, are qualitatively different from understanding how circuit elements combine to produce behavior (for recent discussions, see Krakauer et al., 2017; Fregnac, 2017).

Consider the following admittedly crude example. Suppose an alien species is studying how our automobiles work by using an advanced form of technology. Unbeknownst to us, they are able to measure and manipulate our cars while we drive. Suppose they are able to pull the throttle wire linked to the gas pedal, thereby accelerating a car. By doing so enough times, they establish a causal link between the throttle wire and the car's speed (and publish their results in one of their top journals). Obviously, although they can now deduce that the gas pedal plays an important role in the car's movement, their understanding of automobile function is increased only minimally. For one, they have no idea that the system controls air inflow and therefore controls fuel injection into the engine[5]; not to mention the principles of the combustion engine.

Now, consider the following study, where optogenetic stimulation was used to activate neurons in the ventral tegmental area (VTA) of the midbrain of mice (Pascoli et al., 2018). As in the classic self-stimulation study by Olds and Milner (1954), mice learned to press a lever, in this case for optogenetic-driven enhancement of VTA activity. After two weeks of training, upon lever pressing, mice now received a brief electric shock in addition to increased VTA activation. Approximately 60% of the mice persisted in lever pressing impulsively, despite being administered the shock. In addition, the authors found that a pathway from the orbitofrontal cortex to the dorsal striatum (in the forebrain) affected the behavior of the mice; for example, optogenetic inhibition of the pathway made them stop lever pressing. Compulsive behavior (lever pressing) could be suppressed or induced by decreasing or increasing, respectively, the strength of this neural connection. Remarkable as these results may be (they were published in *Nature* as a full-length article), how much closer are we to understanding "compulsive behaviors?" For one, the VTA neurons stimulated by lever pressing do not connect directly with either the orbitofrontal cortex or the dorsal striatum. Clearly, a multisynaptic circuit involving the regions studied must be involved (Keiflin and Janak, 2015), and the mechanisms of action remain unknown. The discovery of the simple causal link between the orbitofrontal cortex and the dorsal striatum via sophisticated techniques, intriguing as it is, leads to an ostensible shift of the goalpost but leaves us minimally closer to understanding the circuit. The objective here was not to critique this particular study, but instead to highlight shortcomings of the causal-mechanistic, interventionist approach currently widespread in neuroscience (see also Krakauer et al., 2017).

---

[5] It is assumed that an older car is being studied, and not a newer drive-by-wire model.

## 5. Transient brain dynamics

The upshot is that simple ways of reasoning about causation are inadequate when unravelling the workings of a complex system such as the brain. Instead of focusing on causation as the inherent goal of explanations in neuroscience, a fruitful research avenue is to develop formal tools that describe the *dynamic multivariate structure* of brain data. In other words, one is interested in describing the joint state of a set of brain regions, and how this joint state evolves temporally. A major goal is then to work out how groups of regions *dynamically* coalesce into coherent functional units and how they dissolve when their assembly is no longer needed to meet processing demands.

Consider a system of neurons, neuronal populations, or brain regions, which is characterized by their activation strengths as a function of time: $x_1(t), x_2(t), \cdots, x_n(t)$. The vector $x$ describes the current joint state of the system (when evaluated at a given time, *t*), and $x(t)$ describes how this joint state evolves through time. A popular approach to thinking about brain dynamics was based on the notion of steady-state attractors, in which activity levels would converge to equilibrium (for at least some period of time). For example, when started at state $x_I$, the system would evolve temporally and settle in state $x_A$, where $x_A$ is the stable state closest to $x_I$ (Cohen and Grossberg, 1983; see also Hopfield, 1982, 1984). In such networks, an input pattern will cause activity changes until it settles into one pattern, the so-called attractor state (here, $x_A$). We can thus say that the input is associated with the properties of the entire, and specific, *attractor* state, which can be viewed as its representation. However, the type of dynamics in "attractor networks" is limited in the sense that the key element is the state into which the network settles (which can be represented formally by, for example, a minimum in an energy function). Importantly, the path taken to reach the attractor state does not matter.

The idea of "computing with attractors" should be contrasted with the one of computing with *transient dynamics* (Rabinovich et al., 2008; Buonomano and Maas, 2009). Transient dynamics do not require waiting for the system to reach equilibrium, and the succession of states visited by the system provide the representation for the event in question. The temporal window considered is arbitrary; for example, 300 ms after an input stimulus, 500 ms prior to movement initiation, or 20 seconds during a mental event. Figure 3A illustrates the idea in the context of recordings from neurons in the antennal lobe of the locust (Broome et al., 2006), showing the succession of states associated with the presentation of two distinct odors when projected onto a lower-dimensional three-dimensional space (those less familiar with this type of plot may benefit from Figure 4A-B). The original measurements were performed in 87 neurons, and the projection here is simply for illustrative purposes (we will discuss the issue of dimensionality further below). Whereas the trajectories might come arbitrarily close at several time points[6], the entire trajectory provides a potentially unique *signature* for the task in question, such that the transients are input specific, and contain information about what initiated them. Furthermore, the trajectories are assumed to be *stable*. Thus, they are resistant to noise in that they are reliable to relatively small variations in initial conditions.

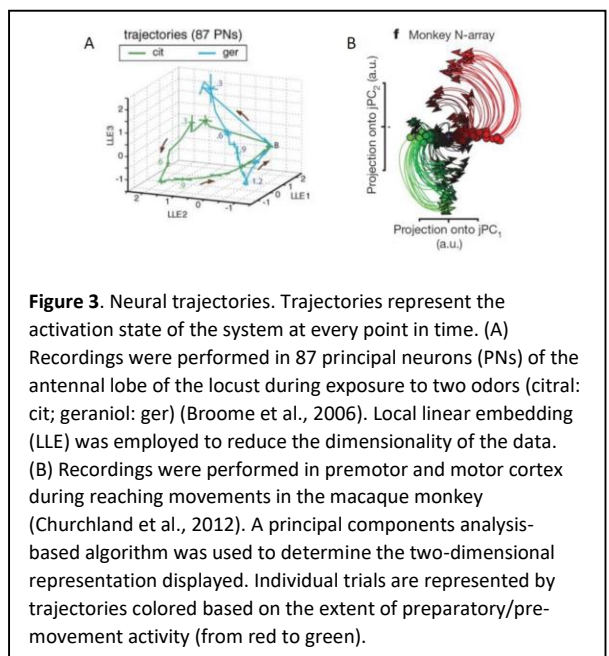

**Figure 3**. Neural trajectories. Trajectories represent the activation state of the system at every point in time. (A) Recordings were performed in 87 principal neurons (PNs) of the antennal lobe of the locust during exposure to two odors (citral: cit; geraniol: ger) (Broome et al., 2006). Local linear embedding (LLE) was employed to reduce the dimensionality of the data. (B) Recordings were performed in premotor and motor cortex during reaching movements in the macaque monkey (Churchland et al., 2012). A principal components analysis-based algorithm was used to determine the two-dimensional representation displayed. Individual trials are represented by trajectories colored based on the extent of preparatory/pre-movement activity (from red to green).

Thinking in terms of trajectories moves the emphasis away from a strictly causal interpretation. Instead of, for example, statements such as "$x_1(t)$ causes $x_2(t+1)$," the framework encourages a description that summarizes the temporal evolution of the system of interest. Experimentally, a central goal then becomes estimating trajectories robustly from available data. At this point, computational models can be tested against the data, or possibly developed to explain the data. In other words, what kind of system, and what kind of interactions between system elements – what mechanisms – generate similar trajectories, given similar inputs and conditions?

---

[6] The issue of the proximity of trajectories will depend on the dimensionality of the system in question (which is usually unknown) and the dimensionality of the space where data are being considered (say, after dimensionality reduction). Naturally, points projected onto a lower-dimensional representation might be closer than in the original higher-dimensional space.

## 5.1 Trajectories during threat processing

Consider a functional MRI participant experiencing alternating "safe" and "threat" blocks, where they lie passively in the former, while they may experience mild shocks during the latter. At the onset of threat blocks, brain regions of the so-called salience network (including the anterior insula and medial PFC) would be expected to respond vigorously compared to the period prior to the block transition (Menon and Uddin, 2010). Regions of the salience network would also respond to the onset of the "safe" block, but suppose that these responses are less vigorous than during the transition to threat. In terms of evoked responses, this scenario can be illustrated as in Figure 4A. For two hypothetical regions, if we diagram the temporal evolution of the responses during the block transitions, the trajectories for the two conditions can be illustrated as in Figure 4B; the state-space plot describes the activity levels $(x_1(t), x_2(t))$. Now, suppose that a group of high-anxious individuals is investigated and that they partially generalize the aversiveness experienced during threat blocks to safe ones; that is, they treat every block onset as a potential transition into a threat condition.[7] In this case, the trajectory observed during safe periods would look more like threat trajectories, and the more so for individuals with higher levels of anxiety (Figure 4E).

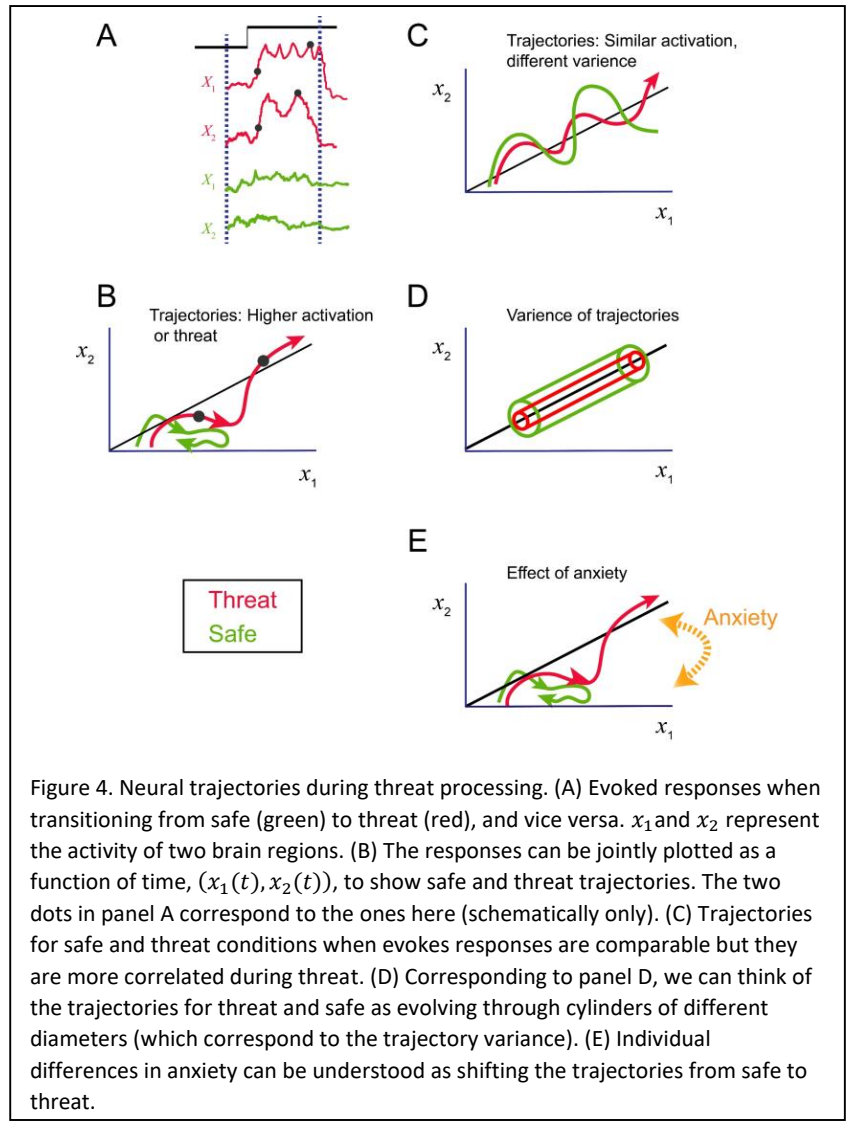

Figure 4. Neural trajectories during threat processing. (A) Evoked responses when transitioning from safe (green) to threat (red), and vice versa. $x_1$ and $x_2$ represent the activity of two brain regions. (B) The responses can be jointly plotted as a function of time, $(x_1(t), x_2(t))$, to show safe and threat trajectories. The two dots in panel A correspond to the ones here (schematically only). (C) Trajectories for safe and threat conditions when evokes responses are comparable but they are more correlated during threat. (D) Corresponding to panel D, we can think of the trajectories for threat and safe as evolving through cylinders of different diameters (which correspond to the trajectory variance). (E) Individual differences in anxiety can be understood as shifting the trajectories from safe to threat.

In an actual functional MRI study that we performed (McMenamin et al., 2014), we found that, somewhat surprisingly, responses evoked when transitioning into safe and threat blocks were rather comparable in magnitude. Presumably, both safe and threat blocks were motivationally significant, thus evoking similar salience-related responses; perhaps, in the context of encountering threat periods, safe periods are quite noteworthy. However, although evoked responses were comparable, signals were more cohesive during threat relative to safe; that is, transitions to threat blocks were associated with evoked responses that were more correlated (for a given pair of regions in the salience network). The respective trajectories for our experiment thus can be illustrated as in Figure 4C (the trajectory linked to threat stays closer to the diagonal ($x_1 = x_2$) than the one linked to safe). And if we consider multiple trials, the trajectories during threat will remain in a part of the space closer to the diagonal (Figure 4D).

## 6. Dimensionality reduction of neural measurements

Neuronal data are inherently high dimensional. Consider, for example, the simultaneous recordings across $10$-$10^2$ locations in electrophysiological grids, $10^2$ sensors with MEG/EEG, $10^2$-$10^3$ neurons with calcium imaging, or the $10^4$-$10^5$ spatial locations with functional MRI. Is it possible that the information across, say, hundreds of measurements could be captured in fewer dimensions without substantial loss of information? Of course, techniques such as principal components analysis are commonplace in data analysis (and can be used, for example, for noise reduction). However, aside from practical concerns, understanding the dimensionality of the data is also important conceptually. For example, it may help uncover relationships that are not apparent in higher dimensions, thus helping to elucidate the mapping from structure to function. In particular, a parsimonious description of the data may uncover stronger relationships with experimentally manipulated variables or other behaviorally relevant variables (see also Santhanam et al., 2009). In addition, the number of dimensions of a dynamical system is an enormously important topic in mathematics (Packard et al., 1980; Takens, 1981; Sauer et al., 1991).

---

[7] High-anxious individuals generalize conditions associated with conditioned fear, for example (see Lissek et al., 2008).

One of the most studied systems in terms of temporal trajectories involves odor processing in invertebrates. In the locust, odors generate distributed responses across the antennal lobe, and such responses evolve in an odor-specific manner (Broome et al., 2006). In Figure 3A, the lower-dimensional representation was obtained by nonlinear dimensionality reduction (Roweis and Saul, 2000). While the dimensionality reduction technique applied was somewhat arbitrary, it helped the investigators gain insight into the following theoretical question: what happens when one odor is being experienced and a second one is presented? One possibility is that the system would "reset," namely responses would return to baseline, then start to evolve in the direction of the new odor (Broome et al., 2006). An alternative possibility would be for the first trajectory (the one associated with the first odor) to deviate from its ongoing evolution and progress along a path corresponding to the mixture of the two odors. Based on the trajectories observed under these experimental conditions, Broome and colleagues were able to rule out the first possibility, while obtaining some support for the second. Together, dimensionality reduction helped uncover mechanisms that would be potentially hard to derive in higher dimensions.

Neuronal dynamics has been investigated in nonhuman primates, too. In one study, Churchland and colleagues (2012) recorded responses in motor and premotor cortex as monkeys performed reaching movements. Data from 50-200 recordings were projected onto two dimensions, revealing a rotational structure to neural trajectories (Figure 3B). Their analysis uncovered processes at the level of the population of neurons, according to which preparatory activity (that is, prior to movement initiation) sets the initial state of a dynamical process that unfolds during movement execution. More generally, the authors proposed that motor cortex expresses a dynamical system that generates and controls movements, and that can be expressed as

$$\frac{d\boldsymbol{r}}{dt} = f\big(\boldsymbol{r}(t)\big) + \boldsymbol{u}(t)$$

where $\boldsymbol{r}$ is a vector describing the firing rate of all neurons (the population response or neural state), $f$ is an unknown function, and $\boldsymbol{u}$ is an external input. As in the example of the locust data, dimensionality reduction helped unearth processes that would not have been evident in higher dimensions.

**7 Geometry of the underlying neural space**

If a neural dataset is acquired in a high-dimensional space and subsequently reduced to a lower dimensionality, what should be the geometry of this space? For simplicity, the original high-dimensional space is frequently, if implicitly, considered Euclidean. But given that not all information can be preserved in fewer dimensions, the question of the nature of the lower dimensionality comes to the fore. For example, in the case of the locust data, a local linear embedding algorithm was employed that attempts to capture information about global geometry in fewer dimensions (by collectively analyzing overlapping local neighborhoods; Roweis and Saul, 2000). In the case of the monkey data, a PCA-based method was applied.

We could follow a similar approach with the functional MRI data of safe and threat periods discussed in Section 5.1 (McMenamin et al., 2014). The study considered 51 brain regions of the so-called salience, executive, and task-negative (also called "default") networks, in addition to the amygdala and the bed nucleus of the stria terminalis (the latter two are particularly important during threat-related processing). We performed dimensionality reduction with the local linear embedding algorithm (Roweis and Saul, 2000) and plotted the mean trajectories for the two conditions, together with an indication of their variance (Figure 5). The two trajectories initially overlap but are quite distinct overall.

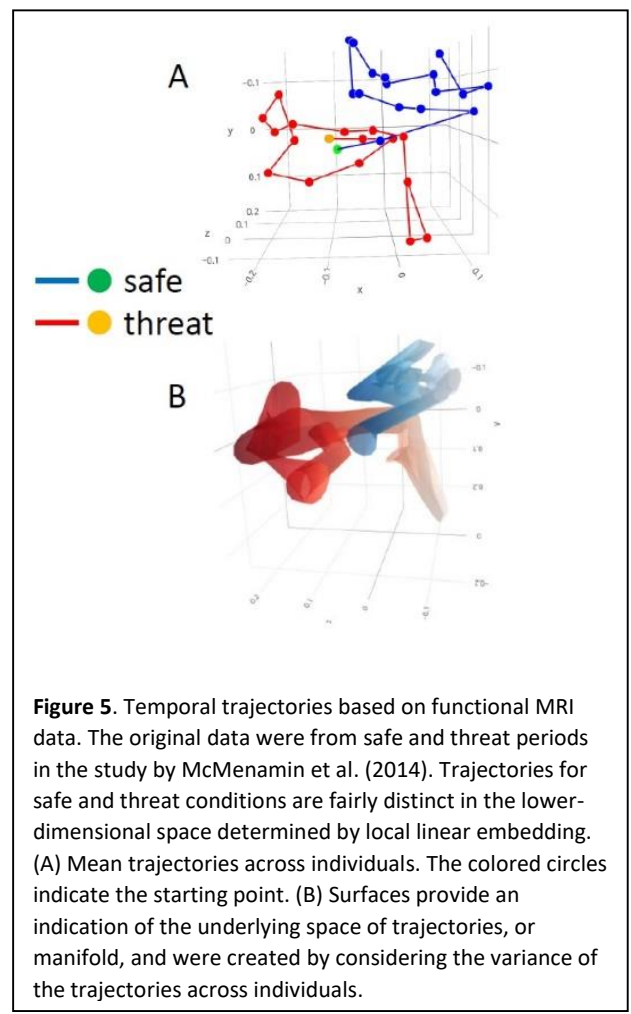

**Figure 5**. Temporal trajectories based on functional MRI data. The original data were from safe and threat periods in the study by McMenamin et al. (2014). Trajectories for safe and threat conditions are fairly distinct in the lower-dimensional space determined by local linear embedding. (A) Mean trajectories across individuals. The colored circles indicate the starting point. (B) Surfaces provide an indication of the underlying space of trajectories, or manifold, and were created by considering the variance of the trajectories across individuals.

More generally, the geometry of the underlying neural space will depend on a combination of the properties of the data and the task condition of interest. We propose the following *neural-dynamics space hypothesis*: behaviors can be described via (they are associated with) classes of trajectories within specific neural spaces (see also Gao et al., 2017). Consider the example of the citral-related trajectory in the locust antennal lobe (Figure 3A). Multiple instances of experiencing this odor are proposed to reside within the surface schematically represented in Figure 6A. This surface defines the space within which trajectories linked with this odor naturally evolve. Such surfaces, which are mathematically called *manifolds*, thus serve as representations of the stimuli, tasks, or conditions in question.

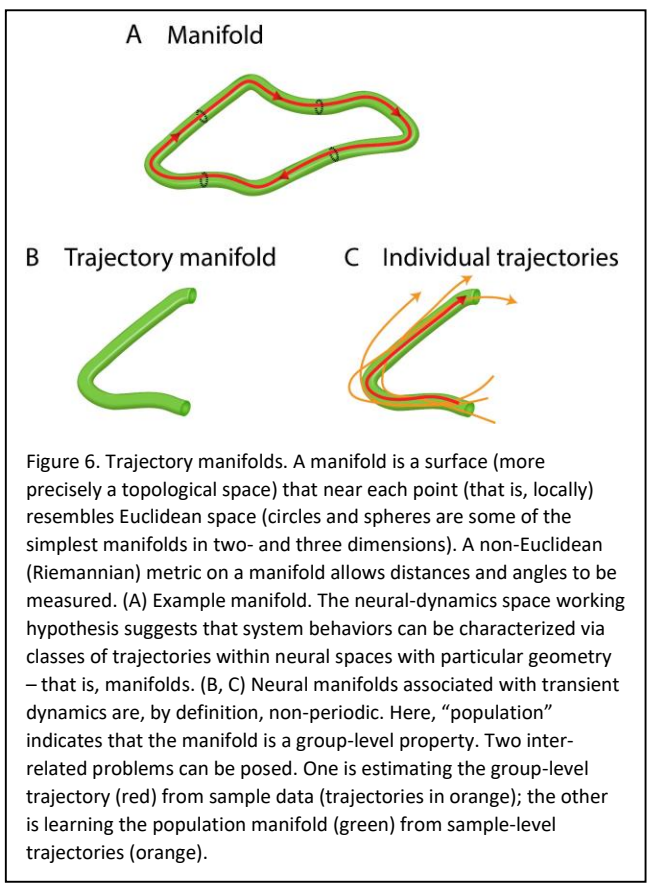

Figure 6. Trajectory manifolds. A manifold is a surface (more precisely a topological space) that near each point (that is, locally) resembles Euclidean space (circles and spheres are some of the simplest manifolds in two- and three dimensions). A non-Euclidean (Riemannian) metric on a manifold allows distances and angles to be measured. (A) Example manifold. The neural-dynamics space working hypothesis suggests that system behaviors can be characterized via classes of trajectories within neural spaces with particular geometry – that is, manifolds. (B, C) Neural manifolds associated with transient dynamics are, by definition, non-periodic. Here, "population" indicates that the manifold is a group-level property. Two inter-related problems can be posed. One is estimating the group-level trajectory (red) from sample data (trajectories in orange); the other is learning the population manifold (green) from sample-level trajectories (orange).

The examples so far assumed the use of an explicit method of dimensionality reduction (the simplest of which is perhaps PCA), a data-driven approach that is suitable in many circumstances. However, knowledge of the problem domain can guide this process, too. In fact, in the case of the Churchland et al. (2012) study, the PCA-based method the authors used was developed to extract rotational information because the authors believed that such coordinate system would be relevant to understanding the topology of neural trajectories in motor cortex during reaching movements[8]. To illustrate the use of domain knowledge, consider a hypothetical study that records multiple cells in each of three distinct brain areas, and suppose that different properties are thought to be important for their function. In this case, one can plot the dynamics of the system in terms of these properties (Figure 7).

---

[8] The rotational structure was not due to primary features of neuronal responses, such as tuning to reach direction (Elsayed and Cunningham, 2017).

In our study of safe and threat periods discussed above (McMenamin et al., 2014), we found that a network-level graph-theory measure called *global efficiency* captured a relevant facet of threat processing. Briefly, efficiency provides a measures of how effectively a network exchanges information (Latora and Marchiori, 2001). In particular, small-world networks are systems that are both locally and globally efficient. Another graph-theory property of interest in our study was node

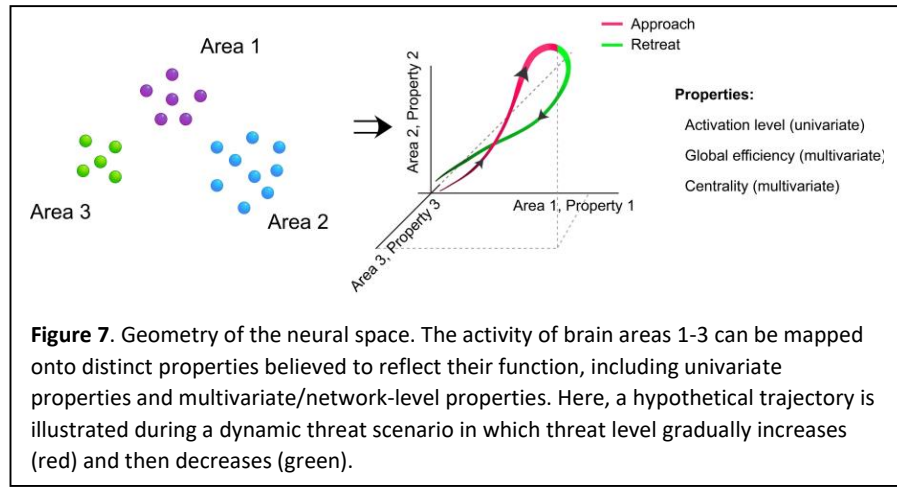

**Figure 7**. Geometry of the neural space. The activity of brain areas 1-3 can be mapped onto distinct properties believed to reflect their function, including univariate properties and multivariate/network-level properties. Here, a hypothetical trajectory is illustrated during a dynamic threat scenario in which threat level gradually increases (red) and then decreases (green).

*centrality*. Increased centrality indicates that a node participates more heavily in the interactions between *other* nodes – that is, they become more of a hub. And, as discussed previously, motivationally significant events, such as blocks transitions, produced stronger responses in regions of the salience network. Accordingly, it could prove informative to project the evolution of the system along these three axes.

Let's apply this idea in the case of a different experimental paradigm. Consider a scenario in which threat is manipulated dynamically. For example, two circles move on the screen in a quasi-random manner and, if they collide, a mild electrical shock is administered to the participant (Myer et al., 2019). Thus, there will be periods of increased anxious anticipation (circles approaching each other) and periods of relative safety (circles retreating from each other). Figure 8 illustrates hypothetical trajectories during approach and retreat in terms of the three dimensions discussed. The overall framework is also fruitful to describe trait- or temperament-like phenotypes. For example, if Figure 8 portrays the situation for a group of low-anxious individuals, for high-anxious individuals one could hypothesize that (*i*) periods of approach would be associated with higher activation of salience-network regions, (*ii*) higher network efficiency, and (*iii*) increased centrality of regions such as the bed nucleus of the stria terminalis and amygdala. Importantly, these properties evolve temporally, as observed experimentally (McMenamin et al., 2014; Najafi et al., 2017; see also Pessoa and McMenamin, 2017).

Dispositional negativity refers to a fundamental dimension of childhood temperament and adult personality and constitutes a prominent risk factor for the development of pediatric and adult anxiety disorders (Hur et al., 2018). Behaviorally, dispositional negativity is associated with threat-related attentional bias and deficits in executive control. Key brain systems proposed to underpin dispositional negativity include the amygdala, as well as the frontoparietal and cingulo-opercular networks (Hur et al., 2018).

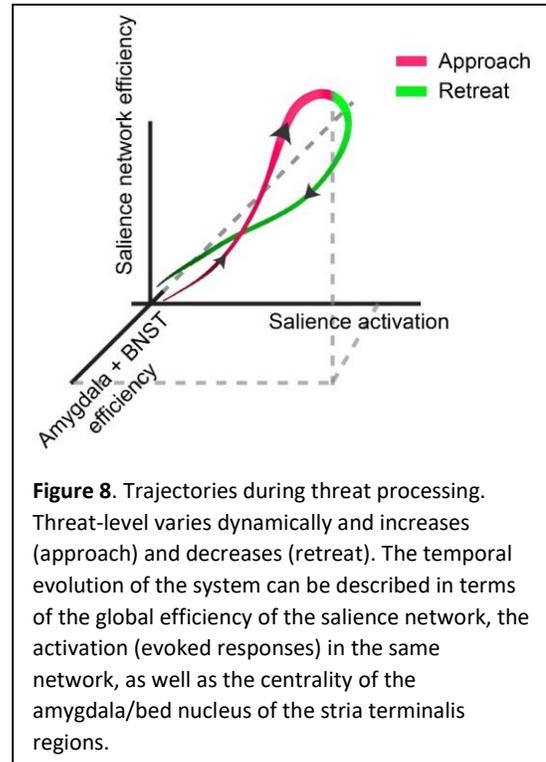

**Figure 8**. Trajectories during threat processing. Threat-level varies dynamically and increases (approach) and decreases (retreat). The temporal evolution of the system can be described in terms of the global efficiency of the salience network, the activation (evoked responses) in the same network, as well as the centrality of the amygdala/bed nucleus of the stria terminalis regions.

One could further test these ideas by investigating experimental conditions involving the performance of cognitively demanding tasks during the presence of threat. In particular, imagine that during the execution of an executive task the threat level is increased from low to high. In terms of neural trajectories, one could hypothesize that there would be a shift in the state-space region occupied by the conditions at hand (Figure 9). In addition, for individuals with higher dispositional negativity two predictions could be made: (*i*) the transition from one region to another would take place faster; and (*ii*) the extent of the change would be greater (that is, the two regions would be farther apart). Irrespective of the potential of these particular predictions to advance the understanding of dispositional negativity, they illustrate how hypotheses can be formulated and tested according to the present ideas. Finally, it also encourages a move away from amygdala-centric proposals that dominate the literature.

**8 Causation in complex systems, again**

In many systems, the relationships between entity-level variables cannot be studied independently of the overall system state. Deyle and Sugihara (2011; see references therein) proposed that such emergence-level view may help explain why many natural systems are so difficult to understand and predict. Building on the mathematical work by Takens (1981) and others, Sugihara et al. (2012) describe a powerful conceptualization that can be called *dynamical systems causation*: two variables are causally linked if they participate in the same dynamical system. In other words, the two variables share a common attractor manifold, such that each variable can identify the state of the other (for example, in the Lorenz attractor). Importantly, this notion can be formally and quantitatively developed (see their convergent cross mapping method). What is more, the proposal is general enough to encompass more traditional views, while also capturing the relationship between variables in many complex systems.

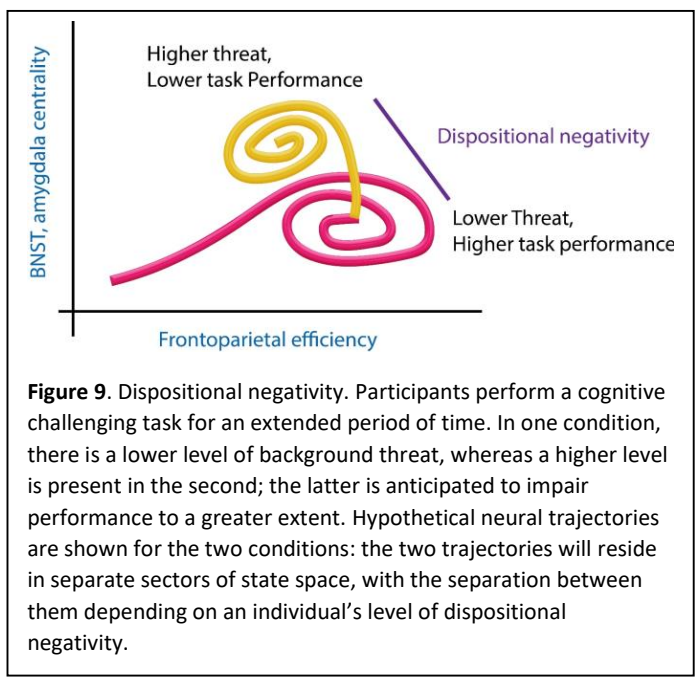

**Figure 9**. Dispositional negativity. Participants perform a cognitive challenging task for an extended period of time. In one condition, there is a lower level of background threat, whereas a higher level is present in the second; the latter is anticipated to impair performance to a greater extent. Hypothetical neural trajectories are shown for the two conditions: the two trajectories will reside in separate sectors of state space, with the separation between them depending on an individual's level of dispositional negativity.

The notion of dynamical systems causation is potentially powerful, but it relies on asymptotic behaviors of the systems in question (such as the attractor manifold). We have emphasized, instead, thinking in terms of transient dynamics, and neural events far from equilibrium and long-term properties. Whereas, conceptually, this does not present a significant impediment, in practice reliably estimating interdependencies may be data-limited. Indeed, these issues have been noted and related methods proposed to ameliorate the problem (Ye and Sugihara, 2016). Furthermore, the problem may be more tractable with group studies in which data from multiple participants is used to recover the underlying manifold. More generally, in group studies, the problem may be posed in terms of estimating a population manifold from individual-level trajectories (Figure 6B-C).

Nevertheless, instead of adopting a single definition/measure of causation, at the current stage of scientific development, it would be beneficial to encourage a plurality of conceptualizations, not least because elucidating complex systems will benefit from multiple vantage points. For example, Mannino and Bressler (2015) propose the notion of *probabilistic causation*: an event does not necessarily determine another event, but rather changes its probability of occurrence. As they state: "A causes B" may be defined as the probability of B given A is greater than the probability of B given that A does *not* occur: P(B|A) > P(B|~A). Their framework is enmeshed with establishing the "ultimate nature" of the brain, namely is it a deterministic or a probabilistic system? Irrespective of this more controversial question, their proposal offers an important way to move beyond outdated models of causes.

It is worth pointing out that the popular framework of sydying causality introduced by Granger (1969) comes with serious limitations (as recognized by Granger himself). In today's terminology, variable $x$ "Granger causes" $y$ if the predictability of $y$ declines when $x$ is removed from the universe of all possible causative variables. However, a key requirement of the model is that of *separability*, such as observed in linear systems. Thus, information about a causative factor needs to be unique to that variable. In coupled systems like the brain, such assumption is clearly violated.

Overall, the suggestion to embrace multiple conceptualizations of causality reflects the idea that the problem is dauntingly challenging. In this regard, it is just the opposite of what was asserted recently by Mehler and Kording (2018): "causality has a perfectly clean definition." Finally, no treatment of causality is probably comprehensive without considering the work of Pearl (2009).

## 9. Learning dynamics with reservoir computing

Temporal trajectories potentially provide signatures for tasks, conditions, or states. Dimensionality reduction provides a strategy to potentially identify trajectories in lower-dimensional spaces. But can trajectories be learned from neural data? In this section, we describe an approach to learning temporal information that we recently developed in the context of functional MRI data (Venkatesh et al., 2019).

Naturally, when considering the temporal information present in functional MRI data, it is necessary to consider the slow evolution of blood oxygenation responses. Accordingly, dynamics should be understood at a commensurate temporal scale – on the order of a few seconds or typically longer. Fortunately, many mental processes unfold at such time scales, such as the processing of event boundaries (Zacks et al., 2001), a gradually approaching threatening stimulus (Najafi et al., 2017), listening to a narrative (Ferstl et al., 2005), or watching a movie (Hasson et al., 2004).

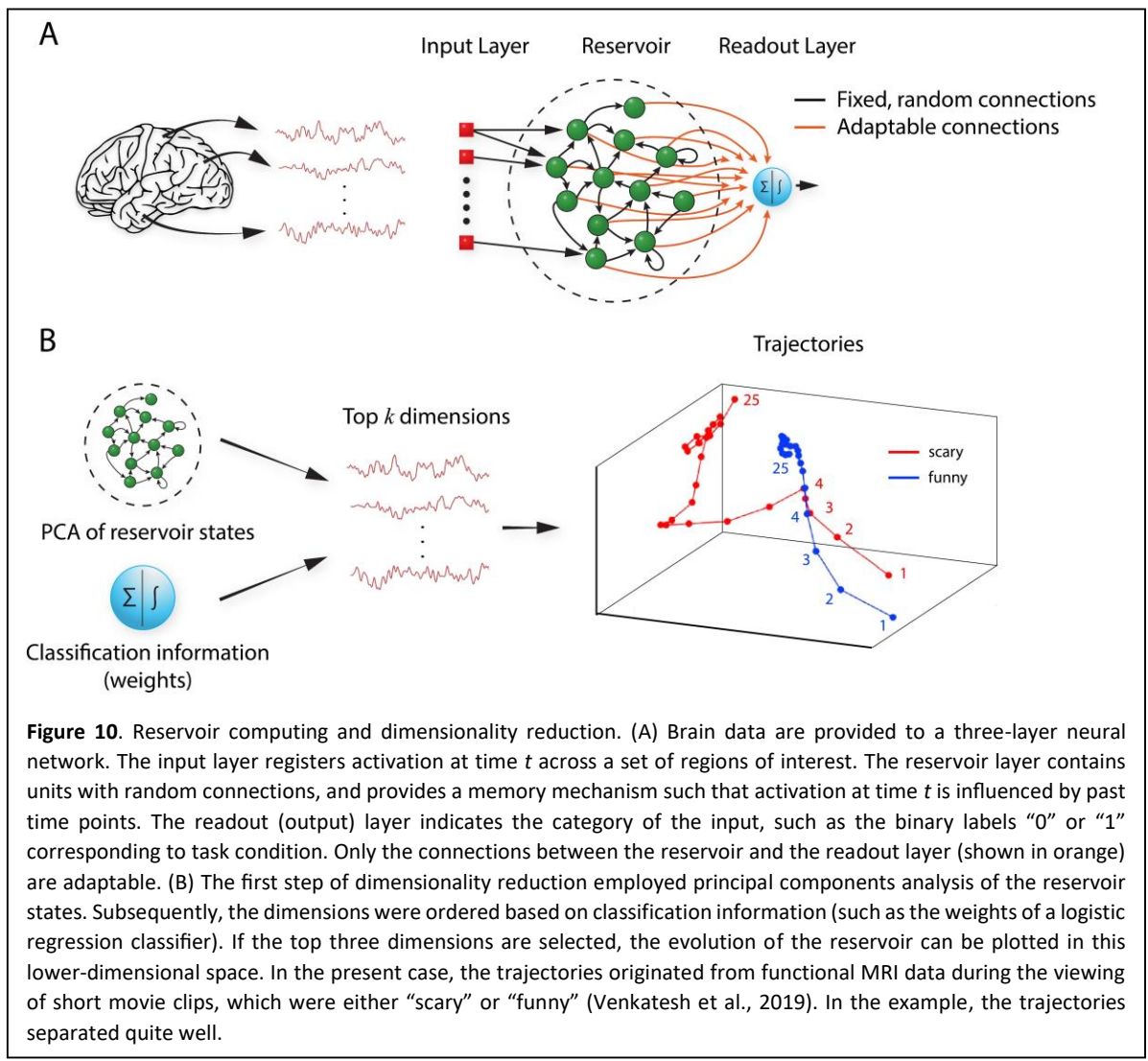

**Figure 10**. Reservoir computing and dimensionality reduction. (A) Brain data are provided to a three-layer neural network. The input layer registers activation at time *t* across a set of regions of interest. The reservoir layer contains units with random connections, and provides a memory mechanism such that activation at time *t* is influenced by past time points. The readout (output) layer indicates the category of the input, such as the binary labels "0" or "1" corresponding to task condition. Only the connections between the reservoir and the readout layer (shown in orange) are adaptable. (B) The first step of dimensionality reduction employed principal components analysis of the reservoir states. Subsequently, the dimensions were ordered based on classification information (such as the weights of a logistic regression classifier). If the top three dimensions are selected, the evolution of the reservoir can be plotted in this lower-dimensional space. In the present case, the trajectories originated from functional MRI data during the viewing of short movie clips, which were either "scary" or "funny" (Venkatesh et al., 2019). In the example, the trajectories separated quite well.

Several machine learning techniques exist that are sensitive to temporal information. Among them, recurrent neural networks (RNNs) have attracted considerable attention (Williams and Zipser, 1989; Pearlmutter, 1989; Horne and Giles, 1995). However, effectively training RNNs can be challenging, particularly without large amounts of data (Pascanu et al., 2013); but for recent developments see (Martens and Sutskever, 2011; Graves et al., 2013). In our study (Venkatesh et al., 2019), we proposed to use *reservoir computing* to study temporal properties of brain data. This class of algorithms, which includes liquid-state machines (Maass et al., 2002), echo-state networks (Jaeger, 2001; Jaeger and Haas, 2004), and related formalisms (Sussillo and Abbott, 2009), includes recurrence (like RNNs) but the learning component is only present in the read-out, or output, layer (Figure 10A). Because of the feedback connections in the reservoir, the architecture has memory properties, that is, its state depends on the current input and past reservoir states. The read-out stage can be one of many simple classifiers, including linear discrimination or logistic regression, thus providing considerable flexibility to the framework. Intuitively, reservoir computing is capable of separating complex stimuli because the reservoir projects the input onto a higher-dimensional space, making it easier to classify them. Of course, this is related to the well-known difficulty of attaining separability in low dimensions, as recognized early on with the use of perceptrons.

Very briefly, the state of the reservoir can be determined as follows:

$$\widetilde{x}(t) = f\left(W^i u(t) + W x(t-1)\right),$$

$$x(t) = (1-\alpha)x(t-1) + \alpha \widetilde{x}(t),$$

where $\widetilde{x}$ is an intermediate state and $x$ is the state of the reservoir with dimensionality $\tau N$, where $\tau$ is a parameter and $N$ is the number of input units; $u$ specifies the input to the system (augmented with a standard bias term of 1). The function $f$ is a sigmoidal function, and $\alpha$ is the forgetting rate parameter. The matrix $W^i$ is the input-to-reservoir matrix and the matrix $W$ specifies the within-reservoir weights, both of which are generated randomly, that is, they are not learned. For more details, see (Jaeger and Haas, 2004; Lukosevicius, 2012).

A central objective of our study was to investigate reservoir computing for the purposes of classifying fMRI data, in particular when temporal structure might be relevant, including both task data and data acquired during movie watching. The latter illustrates the potential of the technique for the analysis of naturalistic conditions, which are an increasing focus of research. One of the conditions we investigated was the so-called "theory of mind" task[9]. Participants watched 20-second clips containing simple geometrical objects (including squares, rectangles, triangles, and circles) that engaged in a potential socially relevant interaction (such as appeareing to initially fight and then make up) that unfolded throughout the duration of the clip. When watching such clips, one has the impression that the potential meaning of the interactions gradually becomes clearer and evolves during the clip. The control condition consisted of same-duration clips using the same geometrical objects following random motion. Could the network classify theory of mind versus random clips, and in what manner was that related to temporal information?

---

[9] From the Human Connectome Project.

Two key parameters determine the memory properties of the reservoir: the forgetting rate, $\alpha$, the and ratio of the number of reservoir-to-input units, $\tau$. Classification accuracy increased as the size of the reservoir increased, and exceeded 85% (which robustly differed from chance levels). We also trained the classifier by randomzing temporal information, namely, by randomly shuffling the data points in a block prior to training, and testing on unperturbed blocks (that is, temporally ordered). In this case, mean classification accuracy was drastically reduced to 56% correct. These and other control analyses indicated that reservoir networks were able to capture some of the temporal dynamics measured by functional MRI.

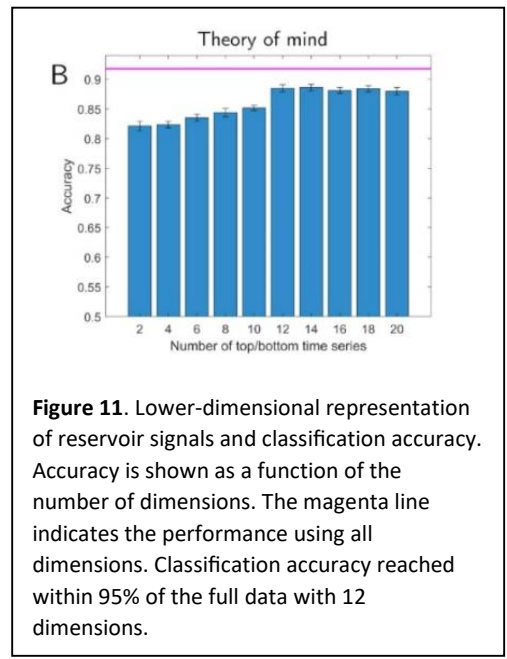

**Figure 11**. Lower-dimensional representation of reservoir signals and classification accuracy. Accuracy is shown as a function of the number of dimensions. The magenta line indicates the performance using all dimensions. Classification accuracy reached within 95% of the full data with 12 dimensions.

We also sought to determine the dimensionality of the reservoir representation capable of classifying task conditions. The original dimensionality of our data was 360, which corresponded to the number of brain regions of interest investigated. As the goal was task classification, we selected dimensions that would contribute the most discriminative information in this regard (Figure 10B). Therefore, we performed PCA on reservoir data (that is, activation of the reservoir layer), and ordered the components based on their contributions to classification (somewhat akin to partial least squares), instead of the variance explained. Figure 11 shows classification accuracy as the number of components was increased from 2 to 20 in steps of two. Remarkably, only ten principal components were required to attain classification at 95% of the level of the full dimensionality of the data. It is noteworthy that these components captured only 7% of the total variance, which should be compared to 70% if one selected components based on the amount of variance explained. Thus, only a small percentage of the original signal variance was informative for classification.

Using just the top three classification-related components allowed good accuracy, as shown in Figure 10A where performance is plotted as a function of time. Accuracy was initially around chance, and increased considerably between time points ~4 to ~8 seconds, eventually surpassing around 80% correct. In terms of the top three dimensions, the trajectories of the social and random conditions initially overlapped, but later became quite distinct (Figure 12B).

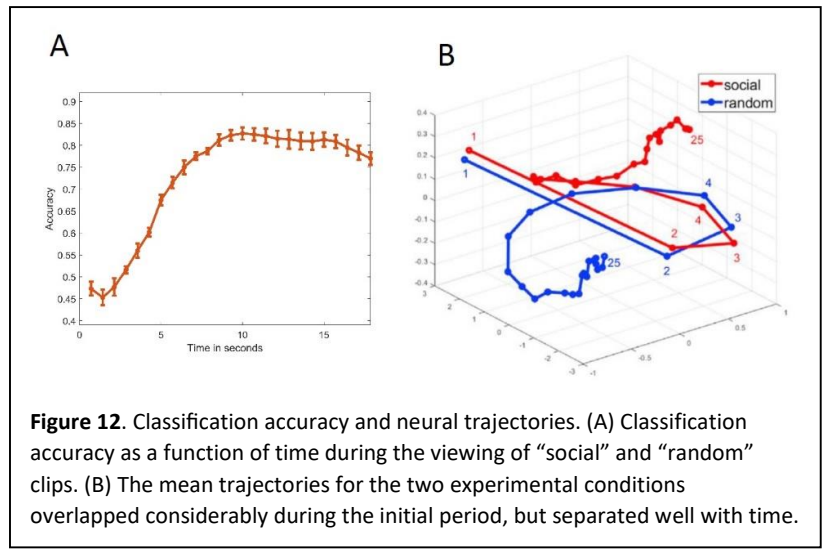

**Figure 12**. Classification accuracy and neural trajectories. (A) Classification accuracy as a function of time during the viewing of "social" and "random" clips. (B) The mean trajectories for the two experimental conditions overlapped considerably during the initial period, but separated well with time.

**10. Conclusions for a science of emotion and cognition**

Neuroscience strives to elucidate the neural underpinnings of interesting behaviors. Modern neuroscience has done so in a preponderantly reductionistic fashion for over a century and a half[10]. I would venture that progress has been stymied by such approach and that the time is ripe for the field to phase-transition into a period when Grossbergian themes come to the fore. By embracing head on the leitmotifs of dynamics, decentralized computation, emergence, selection and competition, and autonomy, a science of the mind-brain can be developed that is built upon a solid foundation of understanding behavior while employing computational and mathematical tools in an integral manner.

At a time when the development of neurotechniques has attained a fever pitch, neuroscience needs to take stock and invest comparable energy in the conceptual and theoretical sides. Otherwise we run the risk of being able to measure every atom in the brain in a theoretical vacuum. Suppose, for that matter, that experimental physicists could measure every atom of a given galaxy. How would that advance understanding if not for a theory of gravitation that took more than 400 years of development? The current obsession in the field with causation is equally problematic. Without theory, "causal" explanations add little to current understanding.

---

[10] We can arbitrarily consider "modern neuroscience" to start with Broca's 1861 clinical report (Broca, 1861).

Ultimately, to explain the cognitive-emotional brain, we need to dissolve boundaries within the brain – perception, cognition, action, etc. – as well as outside the brain, as we bring down the walls between biology, ecology, mathematics, computer science, philosophy, and so on[11]. Let's hope that the body of work by Steve Grossberg can inspire us all in this formidable endeavor.

**Acknowledgements**

The author's research is supported in part by the National Institute of Mental Health (R01 MH071589 and R01 MH112517). I thank Anastasiia Khibovska and Manasij Venkatesh for assistance with figures. I also thank Joyneel Misra and Govinda Surampudi for generating Figure 5.

---

[11] A good example here is quantum physics which has immensely benefited from an intense, if at times strained, exchange between experimental physics, theoretical physics, and philosophy, for example.

**Figure captions**

Figure 1. Fear extinction and structure-function mapping. (A) Fear extinction. (B) Conceptualization of fear extinction in terms of the top-down regulation of the amygdala by the medial prefrontal cortex, with additional variables influencing the process. (C) Schematic representation of the connections between some of the brain regions involved, emphasizing a non-hierarchical view of the processes leading to fear extinction. The descriptors "valence," "regulation," and so on, are not tied to brain areas in any straightforward one-to-one fashion. Abbreviations: CS, conditioned stimulus; MPFC, medial prefrontal cortex; OFC, orbitofrontal cortex.

Figure 2. The three-body problem in Newtonian gravitation. As the problem does not admit to a general mathematical solution, researchers have sought to characterize families of periodic orbits. The figure displays the "yin-yan II" family in two dimensions. The three circles represent the three bodies. Reproduced from Li and Liao (2017). Dynamic plots can be found at http://numericaltank.sjtu.edu.cn/three-body/three-body.htm.

Figure 3. Neural trajectories. Trajectories represent the activation state of the system at every point in time. (A) Recordings were performed in 87 principal neurons (PNs) of the antennal lobe of the locust during exposure to two odors (citral: cit; geraniol: ger) (Broome et al., 2006). Local linear embedding (LLE) was employed to reduce the dimensionality of the data. (B) Recordings were performed in premotor and motor cortex during reaching movements in the macaque monkey (Churchland et al., 2012). A principal components analysis-based algorithm was used to determine the two-dimensional representation displayed. Individual trials are represented by trajectories colored based on the extent of preparatory/pre-movement activity (from red to green).

Figure 4. Neural trajectories during threat processing. (A) Evoked responses when transitioning from safe (green) to threat (red), and vice versa. $x_1$ and $x_2$ represent the activity of two brain regions. (B) The responses can be jointly plotted as a function of time, $(x_1(t), x_2(t))$, to show safe and threat trajectories. The two dots in panel A correspond to the ones here (schematically only). (C) Trajectories for safe and threat conditions when evokes responses are comparable but they are more correlated during threat. (D) Corresponding to panel D, we can think of the trajectories for threat and safe as evolving through cylinders of different diameters (which correspond to the trajectory variance). (E) Individual differences in anxiety can be understood as shifting the trajectories from safe to threat.

Figure 5. Temporal trajectories based on functional MRI data. The original data were from safe and threat periods in the study by McMenamin et al. (2014). Trajectories for safe and threat

conditions are fairly distinct in the lower-dimensional space determined by local linear embedding. (A) Mean trajectories across individuals. The colored circles indicate the starting point. (B) Surfaces provide an indication of the underlying space of trajectories, or manifold, and were created by considering the variance of the trajectories across individuals.

Figure 6. Trajectory manifolds. A manifold is a surface (more precisely a topological space) that near each point (that is, locally) resembles Euclidean space (circles and spheres are some of the simplest manifolds in two- and three dimensions). A non-Euclidean (Riemannian) metric on a manifold allows distances and angles to be measured. (A) Example manifold. The neural-dynamics space working hypothesis suggests that system behaviors can be characterized via classes of trajectories within neural spaces with particular geometry – that is, manifolds. (B, C) Neural manifolds associated with transient dynamics are, by definition, non-periodic. Here, "population" indicates that the manifold is a group-level property. Two inter-related problems can be posed. One is estimating the group-level trajectory (red) from sample data (trajectories in orange); the other is learning the population manifold (green) from sample-level trajectories (orange).

Figure 7. Geometry of the neural space. The activity of brain areas 1-3 can be mapped onto distinct properties believed to reflect their function, including univariate properties and multivariate/network-level properties. Here, a hypothetical trajectory is illustrated during a dynamic threat scenario in which threat level gradually increases (red) and then decreases (green).

Figure 8. Trajectories during threat processing. Threat-level varies dynamically and increases (approach) and decreases (retreat). The temporal evolution of the system can be described in terms of the global efficiency of the salience network, the activation (evoked responses) in the same network, as well as the centrality of the amygdala/bed nucleus of the stria terminalis regions.

Figure 9. Dispositional negativity. Participants perform a cognitive challenging task for an extended period of time. In one condition, there is a lower level of background threat, whereas a higher level is present in the second; the latter is anticipated to impair performance to a greater extent. Hypothetical neural trajectories are shown for the two conditions: the two trajectories will reside in separate sectors of state space, with the separation between them depending on an individual's level of dispositional negativity.

Figure 10. Reservoir computing and dimensionality reduction. (A) Brain data are provided to a three-layer neural network. The input layer registers activation at time *t* across a set of regions of interest. The reservoir layer contains units with random connections, and provides a memory mechanism such that activation at time *t* is influenced by past time points. The readout (output) layer indicates the category of the input, such as the binary labels "0" or "1" corresponding to task condition. Only the connections between the reservoir and the readout layer (shown in orange) are adaptable. (B) The first step of dimensionality reduction employed principal components analysis of the reservoir states. Subsequently, the dimensions were ordered based on classification information (such as the weights of a logistic regression classifier). If the top three dimensions are selected, the evolution of the reservoir can be plotted in this lower-dimensional space. In the present case, the trajectories originated from functional MRI data during the viewing of short movie clips, which were either "scary" or "funny" (Venkatesh et al., 2019). In the example, the trajectories separated quite well.

Figure 11. Lower-dimensional representation of reservoir signals and classification accuracy. Accuracy is shown as a function of the number of dimensions. The magenta line indicates the performance using all dimensions. Classification accuracy reached within 95% of the full data with 12 dimensions.

Figure 12. Classification accuracy and neural trajectories. (A) Classification accuracy as a function of time during the viewing of "social" and "random" clips. (B) The mean trajectories for the two experimental conditions overlapped considerably during the initial period, but separated well with time.

Fox, A. S., Lapate, R. C., Shackman, A. J., & Davidson, R. J. (Eds.). (2018). *The nature of emotion: Fundamental questions*. Oxford University Press.

Frankel, L. (1986). Mutual causation, simultaneity and event description. *Philosophical Studies, 49*(3), 361-372.

Frégnac, Y. (2017). Big data and the industrialization of neuroscience: A safe roadmap for understanding the brain?. *Science, 358*(6362), 470-477.

Fuster, J. M. (2001). The prefrontal cortex—an update: time is of the essence. *Neuron, 30*(2), 319-333.

Gao, P., Trautmann, E., Byron, M. Y., Santhanam, G., Ryu, S., Shenoy, K., & Ganguli, S. (2017). A theory of multineuronal dimensionality, dynamics and measurement. *bioRxiv*, 214262.

Gomez-Marin, A. (2017). Causal circuit explanations of behavior: Are necessity and sufficiency necessary and sufficient?. In *Decoding Neural Circuit Structure and Function* (pp. 283-306). Springer, Cham.

Gomez-Marin, A., Paton, J. J., Kampff, A. R., Costa, R. M., & Mainen, Z. F. (2014). Big behavioral data: psychology, ethology and the foundations of neuroscience. *Nature neuroscience, 17*(11), 1455.

Granger, C. W. (1969). Investigating causal relations by econometric models and cross-spectral methods. *Econometrica: Journal of the Econometric Society*, 424-438.

Graves, A., Mohamed, A.-r., Hinton, G., 2013. Speech recognition with deep recurrent neural networks. In: Acoustics, Speech and Signal Processing (icassp), 2013 IEEE International Conference on. IEEE, pp. 6645–6649.

Grossberg, S. (1964). The theory of embedding fields with applications to psychology and neurophysiology. *Rockefeller Institute for Medical Research*, monograph (451 pp.).

Grossberg, S. (2018). Desirability, availability, credit assignment, category learning, and attention: Cognitive-emotional and working memory dynamics of orbitofrontal, ventrolateral, and dorsolateral prefrontal cortices. *Brain and Neuroscience Advances, 2*, 2398212818772179.

Grossberg, S., & Levine, D. S. (1987). Neural dynamics of attentionally modulated Pavlovian conditioning: blocking, interstimulus interval, and secondary reinforcement. *Applied optics, 26*(23).

Hasson, U., Nir, Y., Levy, I., Fuhrmann, G., Malach, R., 2004. Intersubject synchronization of cortical activity during natural vision. Science 303 (5664), 1634–1640.

Heimer, L., van Hoesen, G. W., Trimble, M., & Zahm, D. S. (2007). *Anatomy of neuropsychiatry: The new anatomy of the basal forebrain and its implications for neuropsychiatric illness*. Burlington, MA: Academic Press.

Herry, C., Ciocchi, S., Senn, V., Demmou, L., Müller, C., & Lüthi, A. (2008). Switching on and off fear by distinct neuronal circuits. *Nature, 454*(7204), 600.
35